\documentclass[11pt,titlepage]{article}
\usepackage{setspace}
\usepackage{amsmath}
\usepackage{amsfonts}
\usepackage{times}
\usepackage{epsfig}
\usepackage{graphicx}
\usepackage{color}
\usepackage{afterpage}
\usepackage[numbers,sort&compress]{natbib}

\begin{document}

\title{Cytoplasmic streaming in plant cells: the role of wall slip}


\author{K. Wolff, D. Marenduzzo, M. E. Cates \\
SUPA, School of Physics and Astronomy, \\
University of Edinburgh, \\
Mayfield Road, \\
Edinburgh EH9 3JZ, Scotland}

\maketitle

\abstract 
{
\vspace*{-20cm}
Accepted (and published online) in J R Soc Interface\\
doi 10.1098/rsif.2011.0868
\vspace*{19cm}

We present a computer simulation study, via lattice Boltzmann simulations, of a microscopic model for cytoplasmic streaming in algal cells such as those of {\it Chara corallina}. We modelled myosin motors tracking along actin lanes as spheres undergoing directed motion along fixed lines. The sphere dimension takes into account the fact that motors drag vesicles or other organelles, and, unlike previous work, we model the boundary close to which the motors move as walls with a finite slip layer. By using realistic parameter values for actin lane and myosin density, as well as for endoplasmic and vacuole viscosity and the slip layer close to the wall, we find that this simplified view, which does not rely on any coupling between motors, cytoplasm and vacuole other than that provided by viscous Stokes flow, is enough to account for the observed magnitude of streaming velocities in intracellular fluid in living plant cells. }


\section*{Introduction}

Within eukaryotic cells, intracellular flows are often unremarkable: typical Reynolds numbers are low, and cells are tiny enough to quench most fluid motion within them. An important exception is provided by {\em cytoplasmic streaming}, which is the name given to the $\sim 80-100$ $\mu$m/s directed flow of cytosol and organelles around large plant and fungal cells~\cite{squires,shimmen_sliding_2007,shimmen_cytoplasmic_2004}~\footnote{Cytoplasmic streaming may also be observed in non-plant cells, for instance in the nematode {\it Caenorhabditis elegans}~\cite{niwayama_pnas_2011}.}. Unlike typical mammalian cells, which are tens of microns in size, the internode stalks of the {\em Chara corallina} algae, for instance, are single cells which are about a millimeter in diameter and several centimeters long. Giant single cells like these need to overcome a highly non-trivial transport problem in order to deliver nutrients and other chemicals throughout their interior. Given such length scales, thermal diffusion does not provide a viable option to move things around quickly enough: for instance, given typical intracellular diffusion coefficients, it would take months even for highly mobile ions to traverse the cell length. Cytoplasmic streaming gives a very efficient alternative, as it sets the whole fluid within the cell in motion, thereby readily advecting proteins and nutrients which happen to be there -- in other words while the Reynolds number may be infinitesimally small, the Peclet number is instead rather large~\cite{van_de_meent_natures_2008,van_de_meent_pnas_2008}. 

Understanding the basic biophysical mechanism underpinning cytoplasmic streaming has been the subject of an ongoing debate at the interface between fluid dynamics and cell biology. To introduce the competing scenarios proposed in the literature, it is useful to recall the geometry of algal cells such as {\em Chara corallina}. The basic components are the ``internodes'' or internode stalks, which as mentioned above are extra-long single cells. Each of these internodes has an approximately cylindrical symmetry, and may be thought of as consisting of two concentric layers: the external one is a micron-thick cellular layer, filled with cytosol (endoplasm) and separated from the large interior, known as vacuole, by a fluid membrane. The vacuole is made up of an aqueous solvent, and contains no internal structure. The inner surface of the external wall of the internode cell is instead covered by actin bundles, which act as conveyor belts along which myosin motors walk. The actin bundles, or lanes, follow helical paths on the cylindrical walls of {\it Chara}, and they are organised into alternating bands having opposite directions of velocity. The bands are separated by a so-called indifferent zone. The hypothesised mechanism for cytoplasmic streaming is that the directed motion of myosin motors on the actin bundle tracks can somehow entrain the cytoplasmic fluid, and that this in turn sets the vacuole into motion by transferring momentum through the fluid membrane separating it from the cytoplasm. According to this view, the myosin traffic provides an effective shear boundary condition on a Stokes equation for the velocity inside the vacuole. The fact that the tracks along which motors move are helicoidal leads to an interesting fluid dynamics problem. Its solution, presented in~\cite{van_de_meent_natures_2008}, includes a secondary flow along the planes perpendicular to the cylindrical axis, which helps nutrient and chemical mixing along the flow gradient direction as well -- mixing is notoriously a hard requirement to fulfil at zero Reynolds number. More recently, direct NMR velocimetry experiments of flow inside the vacuole of single internodal cells have quantitatively validated this analytical solution~\cite{van_de_meent_measurement_2010}.

Earlier experiments on cytoplasmic streaming based on laser light scattering experiments on {\it Nitella}~\cite{mustacich_observation_1974,mustacich_study_1976} and {\it Chara}~\cite{sattelle_cytoplasmic_1976} cells showed that the fluid velocity distribution is quite narrow. While Sattelle \& Buchan~\cite{sattelle_cytoplasmic_1976} speak of plug-like motion for both {\it Chara} and {\it Nitella}, Mustacich \& Ware~\cite{mustacich_study_1976} underscore the fact  that the peak in the spectrum of scattered light is not sharp enough for the streaming flow of a plug but indicates that most particle velocities are within 10\% of the most likely velocity. Experiments agree that it is unlikely that any particles move at substantially larger velocities than those carried by cytoplasmic streaming -- in other words, the fluid velocity has to be very close to that of the vesicles dragged along by the motors and driving the flow.

The idea that myosin motion can provide an effectively smooth boundary condition for the fluid dynamics inside the vacuole by necessity requires a strong hydrodynamic coupling between actin-myosin motion on the cell surface and fluid flow in the vacuole. Finding exactly what structure can provide this coupling is however a non-trivial task. In an interesting contribution~\cite{nothnagel_hydrodynamic_1982}, Nothnagel and Webb analysed the hydrodynamic feasibility of three models for momentum transfer from myosin to endoplasm and vacuole. Their calculation shows that individual myosin molecules running on the actin tracks are by themselves ineffective in setting the cytosol or the vacuole into motion. A second model considered attachment of myosin to organelles and vesicles in the endoplasm: while this much improves the viscous coupling, the calculations in Ref.~\cite{nothnagel_hydrodynamic_1982} still suggested that if this mechanism was at the origin of cytoplasmic streaming it would require very packed traffic of motors along the actin lanes, which is unlikely in the real system. The final model, favoured by the semianalytical estimates in Ref.~\cite{nothnagel_hydrodynamic_1982}, envisages the existence of an elastic network, or gel, incorporating the moving motors and extending into the endoplasm: in this framework the movement of the motors pulls the network forward in a plug-like fashion and the coupling to the vacuole is readily achieved. 

Subsequent experiments (video-enhanced light microscopy and fast-freezing electron microscopy) further suggested that the myosin motors may actually be attached to the endoplasmic reticulum which then performs a sliding motion over the actin cables~\cite{kachar_mechanism_1988}. In some later papers this picture has been adopted~\cite{yamamoto_chara_2006}, while others are less definite and speak only of cargoes in general~\cite{morimatsu_molecular_2000} or state that the organelles associated with myosins have not been identified~\cite{shimmen_sliding_2007}.

Here we revisit the view that the myosin motors need not be directly associated 
with a network structure but that the high viscosity of the cytoplasm together 
with a thin slip layer are sufficient to make it move with a very flat velocity 
gradient at roughly the same velocity as the active particles. We perform 
lattice Boltzmann simulations of micron-sized spherical vesicles moving close to a planar wall 
(dragged by the motors) and show that, depending on the slip allowed at the 
wall, this simpler picture can explain the uniform streaming profile also for 
realistic densities of spherical particles. The effect of the endoplasmic 
reticulum and all other cell contents is incorporated into the high viscosity of 
the cytoplasm, but our results suggest that it is not necessary to invoke a 
physical tethering of the motors to any network in order to explain the known phenomenology
of cytoplasm streaming.

\section*{Models and methods}

\subsection*{Lattice Boltzmann simulations of colloidal spheres in a fluid}

\begin{figure}[t!]
    \begin{center}
    \includegraphics[width=0.5\textwidth]{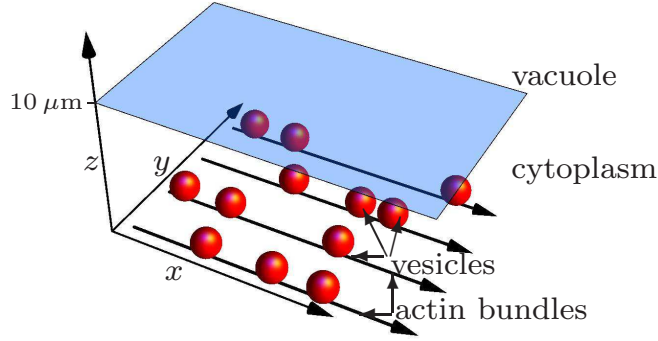}
    \end{center}
    \caption{%
	(Online version in colour.)
        Cartoon illustration of spherical vesicles moving through the cytoplasm 
	(dragged by molecular motors not shown). The cartoon is roughly to scale
	with vesicle diameters 1\,$\mu$m, height of the cytoplasm 10\,$\mu$m and
	thickness of actin bundles 100\,nm. The light blue tonoplast membrane 
	separates cytoplasm and vacuole which extends beyond the figure. In the
	simulations, the motors and the membrane are not modelled explicitly.
     }%
	\label{fig:cartoon}
\end{figure}

Rather than address the cylindrical geometry suggested by the {\it Chara} system, we address here a simpler, planar model (Fig.~\ref{fig:cartoon}). This allows us to address the issues of principle outlined above while avoiding the numerical difficulties of a curved geometry. Moreover, as detailed below, the physics of interest to us concerns the cytosol and only a thin region of the vacuole containing the endoplasm-vacuole interface: this region is almost planar. We consider a slab of material made up of two layers with different viscosities, $\eta_{I}$ and $\eta_{II} > \eta_{I}$, corresponding to respectively the vacuole and the endoplasm, modelled here as a Newtonian fluid with high viscosity, as in the semianalytical treatment in Ref.~\cite{nothnagel_hydrodynamic_1982}.
In each of the layers, the fluid obeys the Navier-Stokes equation,
\begin{equation}\label{NS}
\rho \left(\partial_t+u_{\beta}\partial_{\beta}\right)u_{\alpha}=-\partial_{\alpha}p+\partial_{\beta}\left(\partial_{\alpha}u_{\beta}+\eta_i\partial_{\beta}u_{\alpha}\right)
\end{equation}
where $\rho$ denotes fluid density, ${\mathbf u}$ fluid velocity, $p$ pressure,
$\eta_i$ is the fluid viscosity, and
$i=I,II$ labels the layer under consideration. In our formalism Greek letters denote Cartesian components and summation over repeated indices is implied. In order to solve Eq.~\ref{NS}, we employ our Ludwig code, which is based on a lattice Boltzmann (LB) algorithm (see Ref.~\cite{ludwig} for details). Briefly, the LB method is based on a set of mesoscopic velocity distribution functions, $f_i({\mathbf x},t)$, which depend on some lattice velocity vectors ${\mathbf e}_i$. In our 3D case, we consider 19 velocity vectors (see  Appendix), this is known as a D3Q19 lattice. The density and velocity fields of the fluid may be recovered as moments of such distribution functions, as follows,
\begin{eqnarray}
\rho({\mathbf x},t) & = & \sum_i f_i({\mathbf x},t)  \\ \nonumber
\rho({\mathbf x},t) u_{\alpha}({\mathbf x},t) & = & \sum_i f_i({\mathbf x},t)e_{i,\alpha} .
\end{eqnarray}
In our LB algorithm, the distribution functions evolve according to the following discretised, or lattice, Boltzmann equation:
\begin{equation}\label{LBE}
f_i({\mathbf x}+{\mathbf e}_i\Delta t,t+\Delta t)=f_i({\mathbf x},t)+
\frac{f_i^{\rm eq}({\mathbf x},t)-f_i({\mathbf x},t)}{\tau}
\end{equation}
where $\Delta t$ is the time step (normally unity in LB), $\tau$ is a relaxation parameter, related to the fluid viscosity $\eta$ via $\eta=\frac{\rho}{3}(\tau-\Delta t/2)$, and $f_i^{eq}$ is a set of equilibrium distribution functions. Note that $\tau$ differs in the vacuole and cytoplasm as the viscosities differ. The macroscopic equations of motion obeyed by $\rho$ and $u_{\alpha}$ may be found by taking moments of Eq.~\ref{LBE} and expanding for small $\Delta t$ -- a formal procedure known as the Chapman-Enskog expansion. The LB method defines some dynamical rules for the distribution functions which correspond to a continuity equation for the fluid density and Eq.~\ref{NS} for its velocity field. It can be shown that this is the case provided the following constraints on the moments of the equilibrium distribution are satisfied:
\begin{eqnarray}
\sum_i f_i^{\rm eq}({\mathbf x},t) & = & \rho({\mathbf x},t) \\ \nonumber
\sum_i f_i^{\rm eq}({\mathbf x},t) e_{i,\alpha}& = & \rho({\mathbf x},t) u_{\alpha}({\mathbf x},t)\\ \nonumber
\sum_i f_i^{\rm eq}({\mathbf x},t) e_{i,\alpha}e_{i,\beta}& = & p({\mathbf x},t)u_{\alpha}({\mathbf x},t)u_{\beta}({\mathbf x},t). 
\end{eqnarray}
At this point, it should be noted that any set of equilibrium distribution functions satisfying these constraints leads to a viable LB algorithm to solve the fluid dynamics problem we are interested in. A popular choice in the literature is to expand the equilibrium distribution functions as a power series in the velocity vectors, with the coefficients being determined, in general in a non-unique way, through the constraints. Note that as we want to model a two-layered fluid, made up by a more viscous endoplasm and a less viscous vacuole, we considered a generalisation of the LB equation given above, to allow for a spatially dependent value of the relaxation constant $\tau$. We found this simple procedure to describe well our two-layered system -- we anticipate this may not hold in more complicated systems where the interface is not planar.

The LB method, in the formulation which we adopt, has been developed to allow consistent coupling between the fluid dynamics of a Newtonian fluid and the dynamics of spherical colloidal particles embedded in it. The coupling is through the well-known method of bounce-back on links, which accurately enforces no-slip boundary conditions for the velocity field on the surface of each particle. Instead of repeating the discussion of these colloidal boundary conditions here, we refer the interested reader to~\cite{ladd,ludwig} for details. In our case, the solid particles represent the motors with their associated cargoes (vesicles and organelles). They are positioned at a fixed distance from the bottom wall, and are subjected to a strong harmonic potential which virtually confines their motion to fixed 1D lines, which represent the actin bundles. 

A noteworthy advantage of the LB methodology we use is that, as it fully handles the boundary conditions for the velocity fields both on the particle surface and on the walls, both near and far field effects are correctly included. 
With respect to e.g. Stokesian dynamics, which would be another perfectly valid choice for the problem at hand, LB requires one to introduce both an inertial term and a finite compressibility: both these terms need to be kept small so that they do not effectively contribute to the physics of our system. On the other hand, LB allows one to reach much larger systems than Stokesian dynamics, at least for the nonuniform geometry addressed here. { Finally, we note that an analytical approximation based on modelling the vesicles as Stokeslets near a wall would not incorporate near-field interactions which are important in our simulations, as particles may come into close contact.}

\subsection*{Boundary conditions for slip and no-slip walls}

In this Section we discuss in some detail how to implement slip boundary conditions at the bottom wall (see geometry in Fig. 1), which is important for our work. Here we outline the main equations which we use in our algorithm, whereas a full derivation is given in the { Appendix}. 

In order to model slip at a solid wall in LB simulations, a partial slip  parameter $p$ can be introduced which intuitively corresponds to the ``fraction of slip'' at the wall. More accurately, $p$ is the fraction of each of the distribution functions which is ``bounced forward'' at the wall, whereas $1-p$ is the fraction which is ``bounced back'' (see {Appendix}). The limit of $p\to 0$ corresponds to the commonly employed no slip boundary condition. The amount of wall slip may also be related to the Knudsen number of the fluid~\cite{sbragaglia_analytical_2005}. In our scheme, we use it as a parameter to be determined, in practice, from the knowledge of the slip length, $u(0)/u'(0)$, which we assume comes from either experiment or theory.

Our choice of a slip wall for the bottom boundary is motivated by the need to model the fluid dynamics in the endoplasm close to the dragged cargo. We assume that below the cargo lies a thin aqueous low-viscosity layer, as the viscosity in the endoplasm derives from the presence of a labile reticular network with mesh size smaller than the distance between sphere and wall, or from macromolecular crowding in the endoplasm. Either way, the network or crowding agents, which are coarse-grained into a high-viscosity fluid for the bulk endoplasm, are depleted in the vicinity of the wall, and this accounts for the thin layer of lower viscosity. If the behaviour of the thin layer itself is not of interest, all it does is to introduce effective slip for 
the main system which can be simulated using the slip parameter $p$.

To make progress, we start by writing down the velocity profile of the fluid as a power series in the distance from the bottom wall $z$, 
\begin{equation}
 u(z)=\left\{\begin{array}{ll}
                \sum_{n=0}^N a_n z^n & \quad 0\leq z\leq\epsilon\\
                \sum_{n=0}^N b_n z^n & \quad \epsilon<z\leq R\nonumber
            \end{array}\right.
\end{equation}
The slip length $u(0)/u'(0)$ can be determined analytically for $\epsilon\ll R$ 
where $R$ is the total system size and $\epsilon$ the thickness of the 
low-viscosity layer. By enforcing no slip at the bottom wall, together with continuity of the stress and velocity fields at $z=\epsilon$, and subsequently taking the limit $\epsilon \to 0$ (see { Appendix}), we obtain the following form for the slip length,
\begin{equation}
 \frac{u(0)}{u'(0)}=\epsilon\left(\frac{\eta_{II}}{\eta_{I}}-1\right)
\label{eq:uoverup}
\end{equation}
to first order in $\epsilon$, where $\eta_{I}$ is the (lower) viscosity of the thin
layer close to the wall and $\eta_{II}$ the (higher) {viscosity in the bulk of the system}. Here we assume that the viscosity of the thin layer is the same as that of the vacuole as both are basically aqueous fluids. This is however only a conceptual simplification and not necessary for the derivation of the boundary condition.

The slip length $u(0)/u'(0)$ and slip parameter $p$ may be shown to be related by the following formula (derived in the {Appendix}):
\begin{equation}
\frac{u(0)}{u'(0)}=\frac{(2\tau-1)}{2}\frac{p}{1-p}=3\eta_{II}\frac{p}{1-p}.
\label{eq:uu'ofp}
\end{equation}
The last two equations also lead to the following explicit formula for $p$, 
\begin{equation}
p=\frac{\epsilon\left(\eta_{II}-\eta_{I}\right)}{3\eta_{I}\eta_{II}+\epsilon\left(\eta_{II}-\eta_{I}\right)}.
\label{eq:pexplicit}
\end{equation}
As demonstrated in the {Appendix}, numerical simulations of simple test cases performed with a given $p$ lead indeed to a flow profile that agrees with our analytical calculation. {In the formulas above, all quantities are given in terms of lattice, or simulation, units. To convert these to physical units, $\epsilon$ has to be multiplied by the lattice spacing $\Delta x$, while $\eta$ should be multiplied by the fluid density $\rho$ and by $\Delta x^2$, and divided by the time step $\Delta t$. In lattice, or simulation units, $\rho=\Delta x=\Delta t=1$.

In physical units we choose $\Delta x=0.4$ $\mu$m, $\Delta t=0.08$ ms, and $\rho \sim 2.5 \times 10^7$ kg/m$^3$. With these values, our vesicle size and vacuole viscosity map to $0.5$$\mu$m and 1cP respectively, as in the experiments. Furthermore a velocity of 0.01 in lattice units, typically used in our simulations, maps to a velocity of $50$ $\mu$m/s. However our chosen density is much higher than that of water.\footnote{This is typical in LB work~\cite{ludwig,oliver}. Because the LB algorithm uses fluid inertia to update the dynamics, the most efficient discretisation uses the highest density compatible with still remaining in the low Reynolds number regime. With our chosen density the Reynolds number associated with the vesicle size never exceeds 0.1.}
}

\subsection*{Choice of parameters}

Simulations were run for colloidal particles of radius $a_0=1.25$ 
lattice units (LU), whereas the size of the cytoplasm was 
varied within a physically meaningful regime. In keeping with the 
parameters of ref.~\cite{nothnagel_hydrodynamic_1982} vesicles are assumed to 
have a radius of 0.5\,$\mu$m, meaning that 1\,LU$=0.4\,\mu$m (see above). The 
 spacing between actin cables (or ``lanes'') was fixed to
$d_\textnormal{l-l}=2\,\mu$m$=5$\,LU.
For computational reasons, it is desirable for 
the distance of particles from the wall to be such that there is 
at least one fluid node between wall and particle surface
(otherwise lubrication forces need to be considered). 
Particle centres are therefore placed at $z_\textnormal{p}=2.35$ LU 
(and they do not extend below $z=1.1$ LU). 
The distance between the wall (mid-grid at $z_\textnormal{w}=0.5$) and the particle surface 
is thus $d_\textnormal{w-p}=0.6$\,LU. This agrees well with 
the actual distance in the biological system: diameters of actin cables are 
between 100\,nm and 200\,nm~\cite{nagai_cytoplasmic_1966} and the size of 
{\it Chara} myosin is estimated to be about 
100\,nm~\cite{morimatsu_molecular_2000}, much larger than typical animal
myosin motors. This results in about 
$d_\textnormal{w-p}=0.2-0.3\,\mu\textnormal{m}=0.5-0.75$\,LU. 
Initial spacing between particles (centre to centre) varies with 
density: a separation $d_\textnormal{p-p}=12.5$\,LU for instance
corresponds to a line density of 0.2 (in units
where 1 corresponds to a solid line of particles within a lane).

We fixed the vacuole viscosity to $\eta_{I}=0.02$ and varied $\eta_{II}$
in line with existing estimates; see Ref.~\cite{nothnagel_hydrodynamic_1982} 
where the authors assumed a ratio $\eta_{II}/\eta_{I}$ between 6 and 250. 
In simulations the vacuole has thickness 75\,LU=30\,$\mu$m. 
While this is much too small as the vacuole thickness can be 
$\sim$ 500\,$\mu$m, the location of the upper boundary has virtually no effect on the velocities on the
endoplasm-vacuole interface, which we are ultimately interested in.
We also note that moving the upper vacuole boundary further away would, if
anything, only increase the cytoplasm's velocity with respect to particle 
velocities.

Finally, we note that here we assume that the viscosity of the thin layer in the cytoplasm which cannot be reached by the crowding agents is the same as that of the vacuole. This results in a slip layer of thickness equal to 220\,nm, when $p=0.9$. The thickness of the low-viscosity layer is not known but is restricted by the particle to wall distance $d_\textnormal{w-p}=0.6$\,LU, since the particles have to move within the high-viscosity region to provide effective drag on the main endoplasm layer.

\section*{Results}

Fig. 1 shows a cartoon of the geometry of our simulations, which we recall consists in a two-layer fluid, modelling the endoplasm-vacuole system, with a slip length boundary condition, Eq.~\ref{eq:uu'ofp}. A 2D carpet of molecular motors with their attached cargoes is arranged along regularly spaced actin cables and each of the motors is further subject to a constant external forcing which drives it along the lane. 

\subsection*{Allowing for a finite wall slip can reproduce the phenomenology of cytoplasmic streaming}

\begin{figure}[ht!]
    \begin{center}
      \includegraphics[width=\textwidth]{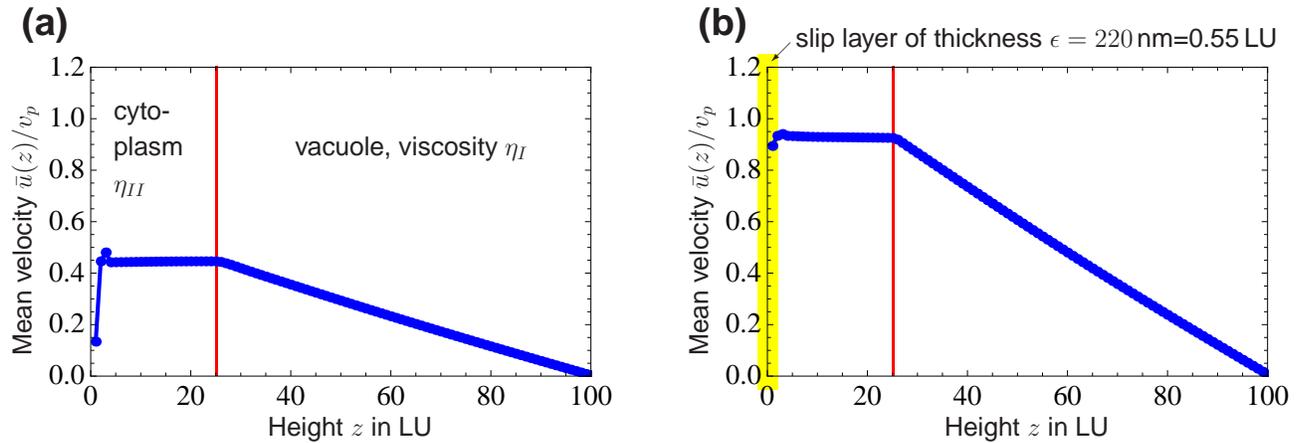}
    \end{center}
    \caption{%
	(Online version in colour.)
        \textbf{(a)} Ratio of average fluid velocity over average particle velocity
	$\bar{u}(z)/v_p$ for a no-slip wall below the vesicles. \textbf{(b)} 
	Partial slip at the bottom wall, corresponding to a layer of lower 
	viscosity and of thickness 220\,nm (or 0.55 LU) close to the wall, can 
	explain the high fluid velocities observed in experiments. Viscosity 
	ratio is $\eta_{II}/\eta_{I}=50$, particle surface coverage $\phi=0.1$ 
	with the distance between actin bundles being $\Delta y=5$\,LU 
	corresponding to 2\,$\mu$m. The height of the cytoplasm is 
	$H_\mathrm{cyto}=25$\,LU=10\,$\mu$m.
     }%
\end{figure}

Fig. 2 shows the mean fluid velocity $u(z)$ as a function of 
$z$, averaged over $x$ and $y$ and normalised by the average velocities of the 
motor carpet -- the latter are recomputed for each simulation as they may vary.
Fig.~2a considers no-slip boundary conditions, $p=0$. Our direct simulations of the 3D flow within the endoplasm suggest that the velocities within the fluid sharply drop to about 50\% of the motor--cargo velocities close to the wall. Interestingly, this is very much in line with the semianalytical calculations presented in Ref.~\cite{nothnagel_hydrodynamic_1982} and relying on more assumptions than our LB simulations. As observed in Ref.~\cite{nothnagel_hydrodynamic_1982}, this is not compatible with experimental measurements which suggest that the velocity distribution within the endoplasm is within 10\% of the motor velocities on the actin cables. The no-slip simulations also show a small peak in $u(z)$, just above the particles. Increasing the size of the particles such a peak disappears, so this is probably a discretization artefact.

Fig. 2b instead shows what happens if we allow for wall slip. As mentioned in the previous section, in the simplest approximation slip may be characterised by a single length scale, henceforth denoted as the slip layer thickness. This is equal to 220 nm in Fig. 2b, which is a reasonable value given the size of the cargoes, and corresponds to a slip parameter $p=0.9$ in our LB simulations. The cytoplasmic flow is now dramatically different and much faster with respect to the no-slip case, and a tracer would move at over 90\% of the mean motor speed, in agreement with the observations. 

\begin{figure}[ht!]
    \begin{center}
      \includegraphics[width=\textwidth]{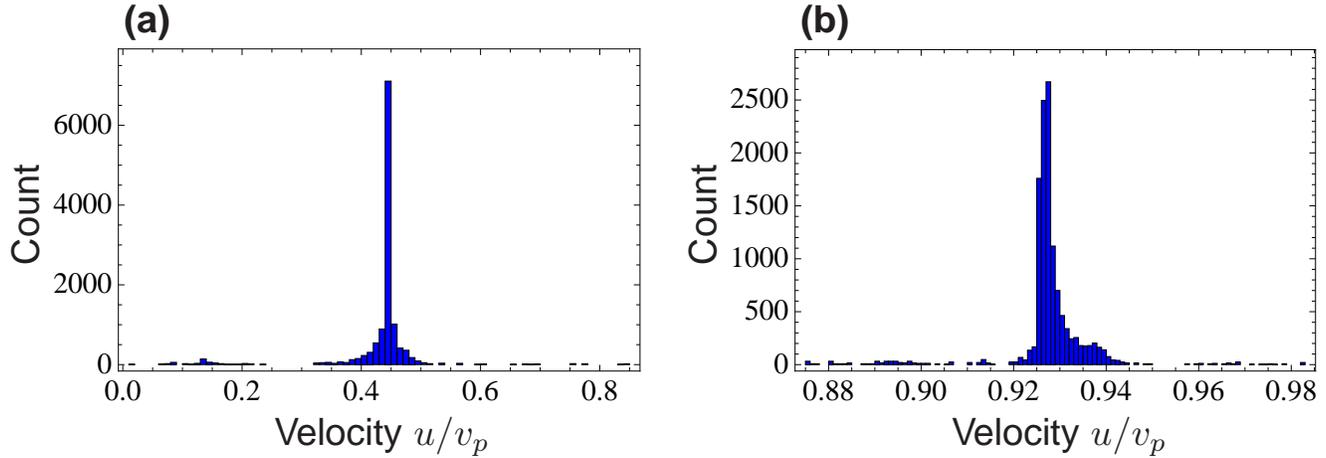}
    \end{center}
    \caption{%
	(Online version in colour.)
      \textbf{(a)} Distribution of the fluid velocities inside the cytoplasm
      for no slip boundary conditions. The fluid velocities are normalised with
      respect to the average vesicle velocity. The distribution is sharply
      peaked, but at less than half the average vesicle velocity.
      \textbf{(b)} Same for a slip parameter $p=0.9$. The distribution is
      now peaked at a value close to the average vesicle velocity in
      agreement with experiments. 
}
\end{figure}

To further compare experiments and simulations, it is useful to go beyond averages and map out the velocity distribution in the cytoplasm. The distribution of velocities of organelles in the cytoplasm is available indirectly from light scattering experiments measuring Doppler shift in algal cells {\it in vivo}: these suggest that the distribution is sharply peaked with only about 10\% variation close to the estimated myosin velocity~\cite{mustacich_observation_1974}. Fig. 3 shows the distributions of the velocities in our simulations, computed on a regular array of points inside the endoplasm, which may be ascribed to tracer organelles within the cell. It can be seen that in the case where slip is allowed our simulated distribution is sharply peaked, and the standard deviation is about 10\% or less of the mean value, in good semiquantitative agreement with experiments. In the no-slip case, the distribution of the endoplasmic velocities is still sharply peaked, but at a much lower velocity than that of the motors, at odds with experiments. 

\subsection*{The flow inside the endoplasm is quasi-one dimensional}

\begin{figure}[ht!]
    \begin{center}
      \includegraphics[width=\textwidth]{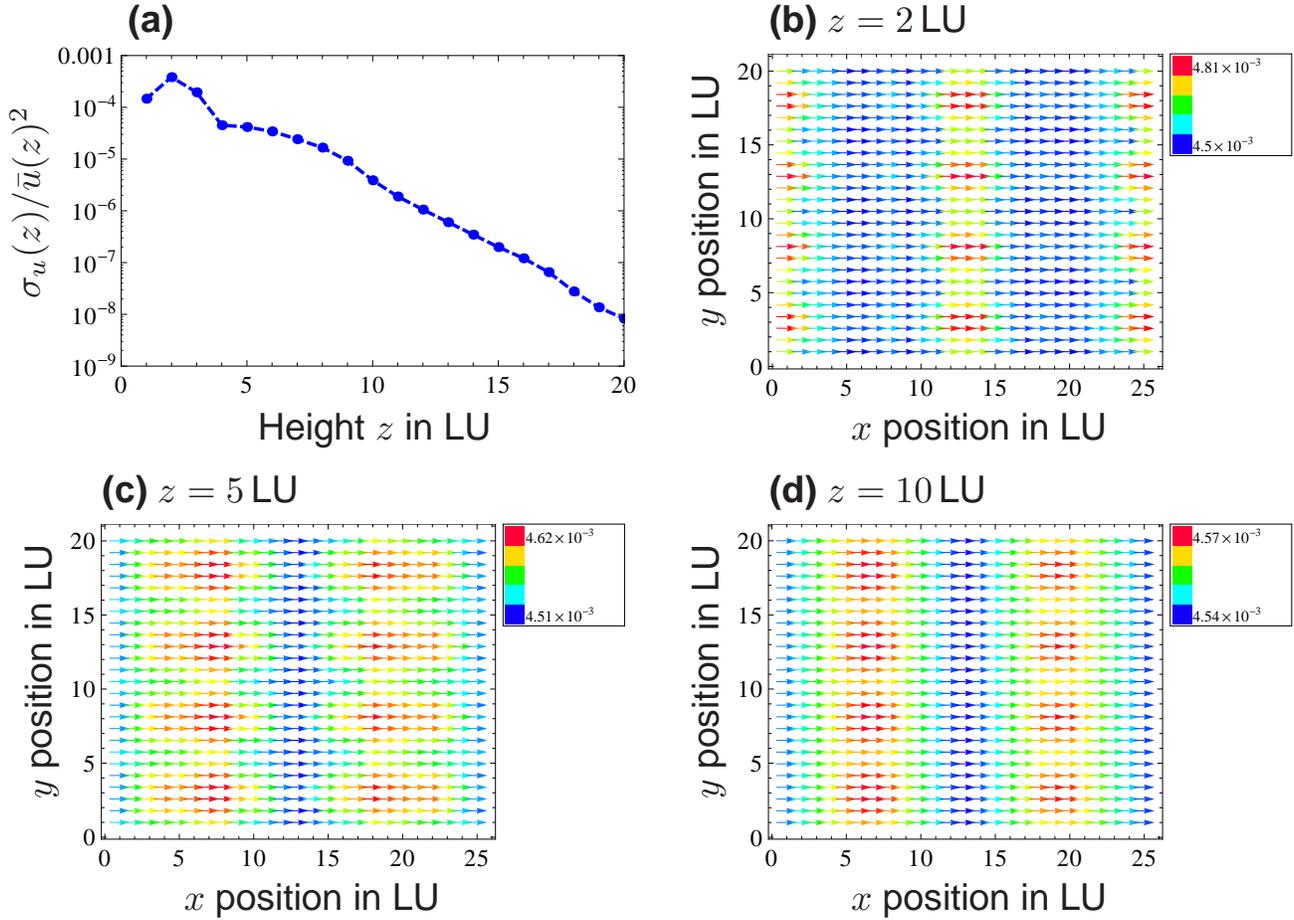}
    \end{center}
    \caption{%
	(Online version in colour.)
        \textbf{(a)} Decay of fluid velocity inhomogeneities with height $z$ as
 	relative variance $\sigma_{u,\mathrm{rel}}=\frac{1}{N-1}\frac{\sum_{i=1}^N 
 	(u_i-\bar{u}(z))^2}{\bar{u}(z)^2}$ where $i$ runs over all fluid nodes
 	in a layer of fixed $z$. Inhomogeneities decay exponentially with $z$. Panels \textbf{(b)} to \textbf{(d)} show 
	velocity maps at slices of fixed $z$ for $z=2$, $z=5$ and $z=10$\,LU. 
        Note the difference in scales for panels \textbf{(b)} to \textbf{(d)}.
     }%
\end{figure}

Our simulations consider a fully 3D flow, with a typical simulation volume equal to $25 \times 20 \times 100$ in lattice units (see {\it Choice of parameters} section for a mapping to real units). As we directly model the two-dimensional carpet of moving motors, together with the interaction with the wall, it is reasonable to expect that close to the motors there will be significant inhomogeneities of the fluid velocity. This may be further increased by any hydrodynamic clustering of the motors. It is then useful to ask how deep inside the endoplasm the correlations and fluctuations in the fluid velocity imparted on the intracellular solvent by the motor dynamics persist, and what the decay of such correlations is. Fig. 4 shows a cut of the fluid velocity profile on different planes, stacked along the velocity gradient direction. 

As can be seen in Fig. 4, close to the particles the flow is indeed inhomogeneous, but surprisingly weakly so. The square root of the velocity variance is only about 1-10\% of the typical fluid velocity even close to the motors; it then decays exponentially with distance and becomes negligible beyond 10 LU away from the walls. In other words, our simulations show that the flow is quasi-1D, corroborating the analysis of Ref.~\cite{nothnagel_hydrodynamic_1982} which assume this geometry at the outset, and also justifying the use of uniform velocity boundary conditions for the vacuole in Ref.~\cite{van_de_meent_natures_2008}.

\subsection*{Experiments constrain the model values of viscosities, motor density and slip layer thickness}

While Figs. 2 and 3 show that the key features of cytoplasmic streaming in plant cells may be reproduced by a reasonable set of values for cytoplasm viscosities, motor geometry, density and slip layer thickness, it should be noted that some of these control parameters are not known with high accuracy from experiments. It is therefore important to assess how our results and conclusions are affected if these are changed, within a biologically relevant range.  In this section we therefore analyse the impact of parameter variation on the fluid velocity within the endoplasm.

\begin{figure}[ht]
\begin{minipage}[t]{0.48\textwidth}
\centering
\includegraphics[width=\textwidth]{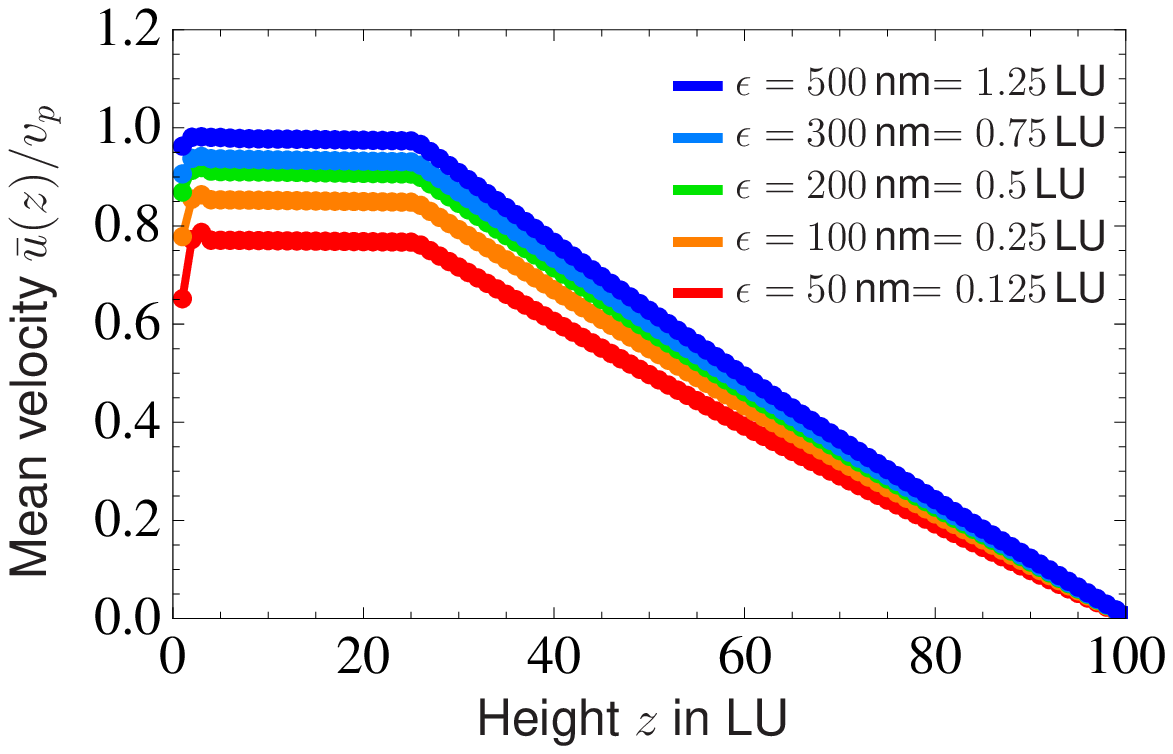}
    \caption{%
	(Online version in colour.)
       	Influence of thickness of slip layer $\epsilon$ with all other 
	parameters ($\eta_{II}/\eta_{I}$, $\phi$, $\Delta y$ and $H_\mathrm{cyto}$)
	held constant. Thicknesses of over 200\,nm for the low-viscosity layer
	are necessary to explain the high $\bar{u}(z)/v_p$ ratio of over 90\%.
     }%
\label{fig:figure5}
\end{minipage}
\hspace{0.5cm}
\begin{minipage}[t]{0.48\textwidth}
\centering
\includegraphics[width=\textwidth]{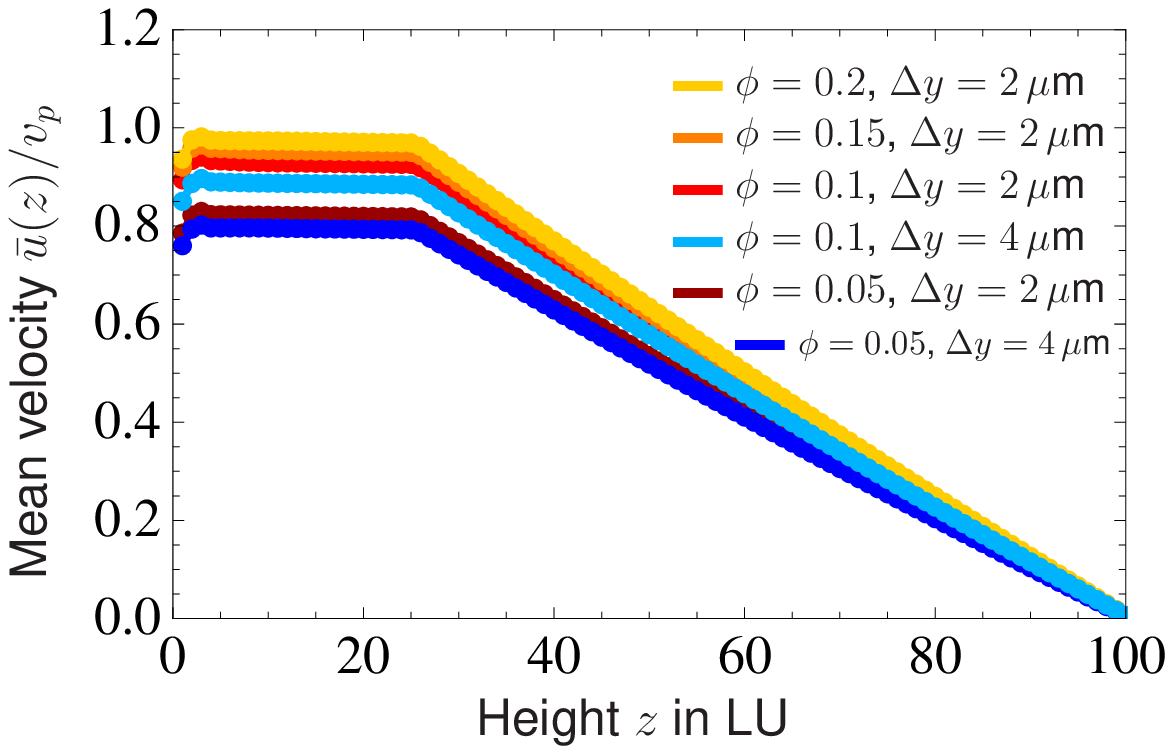}
    \caption{%
	(Online version in colour.)
       	Variation of surface coverage $\phi$ and spacing of actin bundles 
	$\Delta y$. Unsurprisingly, a denser coverage with particles entrains the
	cytoplasm to higher velocities. A surface coverage of 0.1 is necessary 
	to reach high enough velocities, all other parameters being equal.
	The distribution of particles on the surface also plays a role: If the
	particles are dense in the $x$-direction but spaced further apart in the
	$y$-direction the ratio $\bar{u}(z)/v_p$ is lower than if particles
	are denser in $y$ and sparser in $x$ (at constant overall $\phi$). This
	effect is due to channels of slower fluid between actin bundles if they
	are spaced further apart.
     }%
\label{fig:figure6}
\end{minipage}
\end{figure}

We first consider the effect of the thickness of the slip layer, $\epsilon$. Fig. 2b referred to $\epsilon=220$ nm, and Fig. 2a to $\epsilon=0$, with only the former data accounting for a realistic streaming within the cytoplasm. Fig. 5 shows the effect of varying the slip layer, from 50 nm, which is smaller than the size of the motor, to 500 nm, the size of the cargoes. Within this slip layer, we assume that the cytosol is basically an aqueous fluid -- motivated by the view that the larger viscosity inside is essentially due to macromolecular crowding with crowding agents that do not reach up to the wall thus resulting in a lower viscosity close to the wall. The mean velocity shows a saturation behaviour as a function of $\epsilon$, with essentially any value larger than 200 nm enough to account for experimental values. The lower values for slip layer thickness, while geometrically consistent, lead to a velocity of the fluid which is about 80\% of the motor velocity -- this is slightly too low given current measurements, although the corresponding values of $\epsilon$ cannot be definitely discarded.

Fig. 6 shows the effect of changing the density of motors, measured by their surface coverage $\phi=A_\mathrm{occ}/A_\mathrm{tot}$ where $A_\mathrm{occ}$ is the surface area occupied by cargo beads and $A_\mathrm{tot}$ is the maximum area available to them without blocking cargoes on adjacent lanes (i.e. cargoes placed on a rectangular lattice just touching their neighbouring cargoes on the same lane and on adjacent lanes). It is interesting to note that, once the slip layer thickness is large enough, even a surface density equal to 0.1, i.e. 10 times smaller than full coverage, is enough to account for the experimental velocities. Fig. 6 also shows that the distribution of particles on the surface also plays a role. For instance, one may achieve the same surface coverage $\phi=0.1$ by using fewer actin lanes with more myosin motors on each. The largest ratio between fluid and particle velocity is obtained when actin lanes are closely spaced. This effect is due to the creation of channels of slower fluid between actin lanes if the lanes are spaced further apart. 

\begin{figure}[ht]
\begin{minipage}[t]{0.48\textwidth}
\centering
\includegraphics[width=\textwidth]{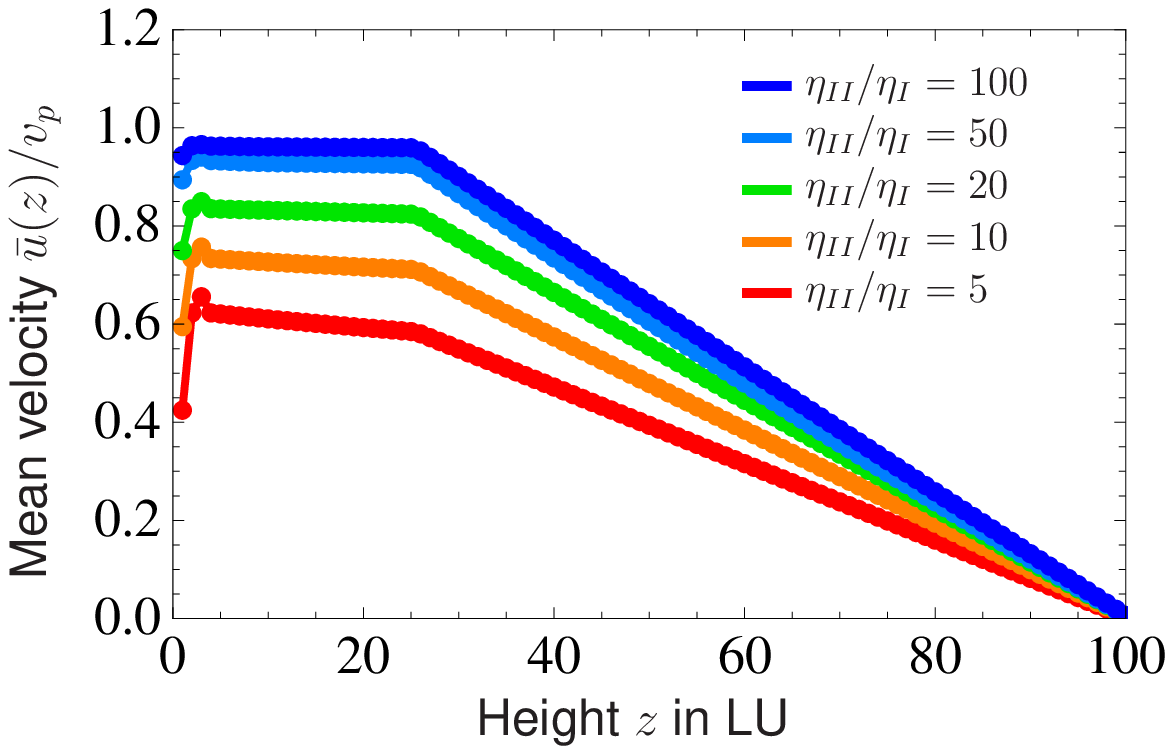}
    \caption{%
	(Online version in colour.)
	Variation of viscosity (ratio of viscosities). A viscosity ratio of
	$\eta_{II}/\eta_{I}$=50 can account for fluid velocities above 90\% of 
	particle velocities while a ratio of $\eta_{II}/\eta_{I}$=20 is not quite
	enough. For velocity ratios larger than $\eta_{II}/\eta_{I}$=100 
	$\bar{u}(z)/v_p$ does not change much anymore.
	Note that for low viscosity ratios the velocity gradient in the 
	cytoplasm also increases such that the velocity at the membrane becomes
	notably different from that at the bottom wall.
     }%
\label{fig:figure7}
\end{minipage}
\hspace{0.5cm}
\begin{minipage}[t]{0.48\textwidth}
\centering
\includegraphics[width=\textwidth]{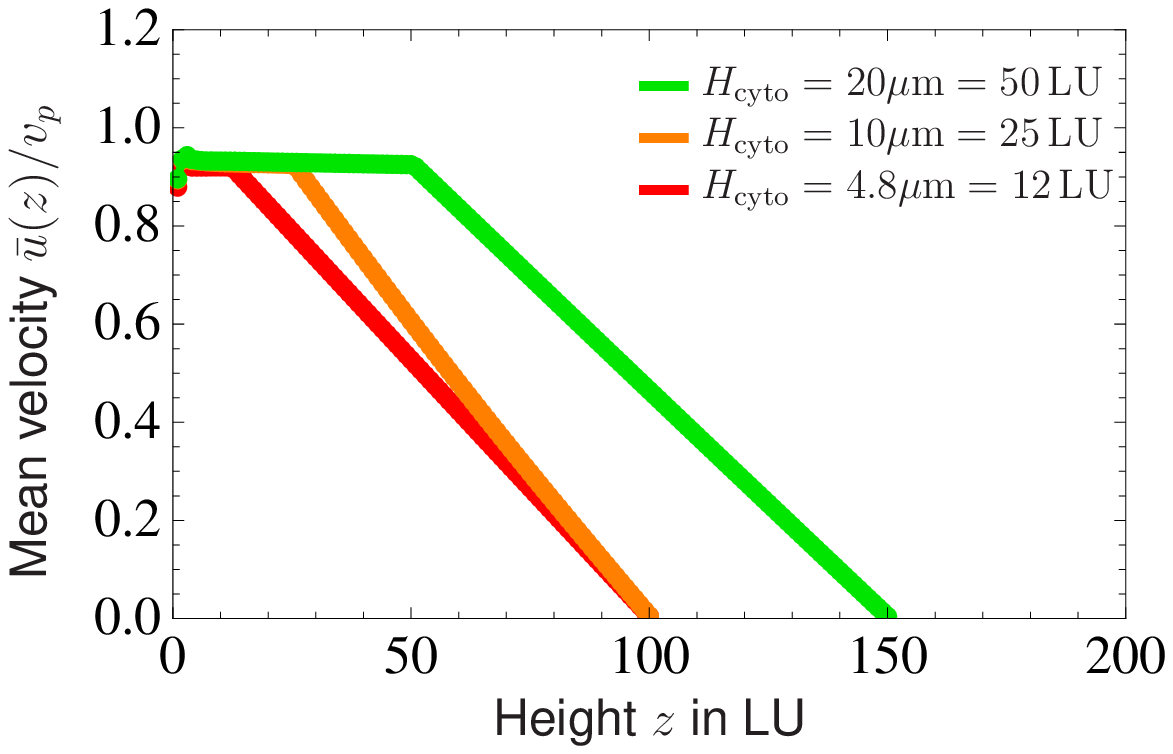}
    \caption{%
	(Online version in colour.)
      Plot of the mean velocity profiles as a function of height along $z$, for 
      different thicknesses of the cytoplasm, $H_\mathrm{cyto}$. One may observe
      that the thickness of the cytoplasm plays no role for the ratio 
      $\bar{u}(z)/v_p$ as long as the top wall is in the far field and 
      the viscosity of the vacuole much lower than that of the cytoplasm. 
      Note that in the curve referring to  $H_\mathrm{cyto}=20$ $\mu$m the
      size of the vacuole has been increased to the minimal one which did
      not affect the fluid dynamics in the cytoplasm.
     }%
\label{fig:figure8}
\end{minipage}
\end{figure}

Finally, Fig. 7 and Fig. 8 show the effects of cytoplasm viscosity and thickness respectively. The viscosity of the cytoplasm has been measured by several authors and values in the range 6-250 cP have been reported, whereas the viscosity of the vacuole may be taken to be close to that of water, 1 cP. The cytoplasm viscosity may actually depend on probe size and this makes it difficult to assess. For our case, as organelles were used in the original experiments to track the flow, it may be more appropriate to use viscosities observed with probes of abou the same size, hence towards the top end of the range quoted above. Fig. 7 shows that the cytoplasm viscosity does have a sizeable effect on the average cytoplasmic streaming velocity: the threshold beyond which realistic streaming may be achieved via cargo dragging is about 20 cP, for which the average fluid velocity is about $\sim 90\%$ of the average motor velocity. Fig. 8 on the other hand shows that the thickness of the cytoplasm plays a comparatively much less important role.

\section*{Discussion and conclusions}

The work we have presented provides a microscopic simulation of cytoplasmic streaming in a geometry relevant to that of plant cells, where there is an endoplasmic layer surrounding a much larger vacuole, and the flow inside the endoplasm is driven by molecular motors which run on actin lanes dragging vesicles. Therefore our calculations can probe the viability of current continuum fluid dynamic theories of cytoplasmic streaming, which often lump the microscopic detail of the motor motion into a shear velocity boundary condition for a purely continuum problem inside the vacuole~\cite{van_de_meent_natures_2008}. 

Our main conclusion is that if motors on the cytoplasm are attached to vesicles or organelles, the drag exerted on the fluid through their motion is enough to lead to an essentially continuum flow within the endoplasm and the vacuole, with average fluid velocity fully consistent with those observed in cytoplasmic streaming. This conclusion holds for a range of realistic parameters, and suggests that there is no need to invoke a solid-like coupling between the motors moving along the subcortical actin tracks and the cytoskeletal network, or endoplasmic reticulum. Such an elastic coupling was deemed necessary in order to account for the experimental velocities in the previous work of Ref.~\cite{nothnagel_hydrodynamic_1982}, and it is therefore useful to point out some key differences between our approaches. Most important is the fact that our work presents direct hydrodynamic simulations of the motion of solid spheres close to a boundary in a two-layered fluid mimicking the endoplasm-vacuole of an algal cell. As such, we fully take into account sphere-sphere and sphere-wall interactions, and crucially also consider slip at the wall. The hydrodynamic treatment in Ref.~\cite{nothnagel_hydrodynamic_1982} on the other hand neglected these interactions, and this is probably at the origin of the quantitative discrepancy between our conclusions (although it should be noted that even the results of Ref.~\cite{nothnagel_hydrodynamic_1982} did not definitely rule out the purely viscous coupling advocated in our work). Indeed, many-body hydrodynamics leads to a collective effect which increases the drag exerted on the fluid by a moving carpet of motors, whereas the wall, which could counteract that effect in principle, does not counteract it in our simulations due to the inclusion of a slip boundary condition.
{ Our approach of introducing slip boundary conditions can likely be generalised to a range of biological solvents whose high effective viscosity is due to macromolecular crowding, so that a layer of lower effective viscosity forms close to walls, where the crowding agents (which are macromolecules and biopolymers) are depleted~\cite{Eisenriegler}.}

It is also important to recognise some remaining limitations of our calculation. Firstly, the interior of the plant cell we wish to model is actually cylindrical, while we have modelled a slab of fluid. This is perhaps not too important as the radius of curvature of the cell is much larger than the thickness of the endoplasm, which is the region of interest, as this is where the coupling between motor motion and vacuolar fluid dynamics occurs. Secondly, we have neglected all details of the intracellular fluid and just modelled the cytoplasm/endoplasm as a Newtonian viscous fluid. In reality, the cytoplasm is non-Newtonian, shear thinning, and it would be interesting to see how a more careful model of the intracellular solvent may affect our results. To this end, we should however make a decision as to whether to model the endoplasm as simply a crowded medium or to include the endoplasmic reticulum as a polymer network. Given the uncertainty with which several of the parameters we use are known, this does not yet seem appropriate and might have to await more detailed experimental study of the rheological propertites of the endoplasm. {In this context it is also interesting to note that in Ref.~\cite{niwayama_pnas_2011}, where the authors studied cytoplasmic streaming in {\it C. elegans}, simulations at constant viscosity proved sufficient to quantitatively reproduce the streaming motion of the (actually non-Newtonian) cytoplasm.} A final simplification of our treatment is that the membrane separating endoplasm and vacuole was computationally treated as an infinitesimally thin interface where the fluid velocity was continuous, whereas in reality this membrane is better described as a viscoelastic membrane.

This work was supported by a DAAD Postdoctoral Fellowship to KW and by EPSRC Grant EP/E030173. MEC holds a Royal Society Professorship.

{
\section*{Appendix}

In this Appendix we give additional information on our numerical framework and 
on the derivation of the formulas determining the partial slip parameter in 
our simulations, see Eq.~\ref{eq:uoverup}-\ref{eq:pexplicit}.

\begin{figure}[ht!]
    \begin{center}
 \includegraphics[width=\textwidth]{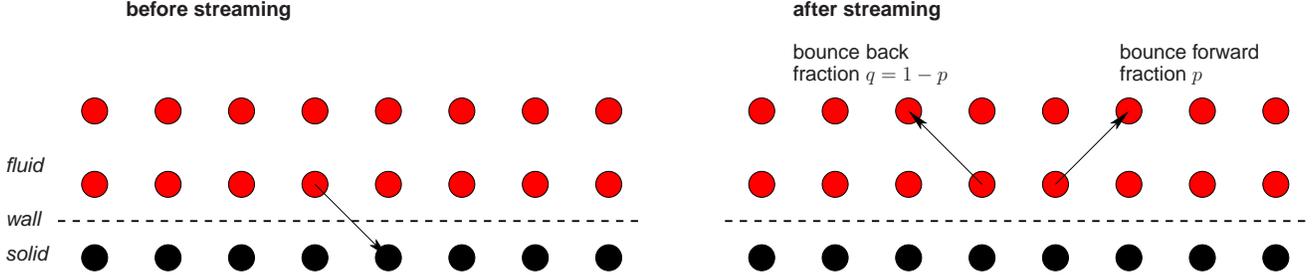}
    \end{center}
 \caption{(Online version in colour.) Velocity populations impinging on the wall are partially reflected 
	forwards (fraction $p$) and partially bounced back (fraction $q=1-p$).}
 \label{fig:slipparameter}
\end{figure}

In order to model slip at a solid wall in lattice Boltzmann (LB) simulations, a partial slip 
parameter $p$ can be introduced which specifies the reflectivity of the wall.
This parameter determines the fraction of a velocity population which is
reflected specularly (or bounced forward) at the wall in an LB streaming step
(see Fig.~\ref{fig:slipparameter}).
Note that $p=0$ corresponds to full no slip boundary conditions for the velocity field at the wall, whereas $p=1$ corresponds to full slip -- we refer to any value of $p$ in between as {\em partial slip}.
The wall reflectivity can be interpreted physically in terms of the Knudsen 
number which enters the simulation via the relaxation time 
$\tau$~\cite{sbragaglia_analytical_2005}. 
Another possibility is to interpret the wall slip in 
a more general way where $p$ is simply an open parameter that can be 
determined if the slip length $u(0)/u'(0)$ is known from experiment or theory.

One particular setting that can be described in this way is that of a 
two-viscosity system where a thin low-viscosity layer is situated close to the 
wall and a bulk fluid of higher viscosity beyond. If the behaviour of the thin 
layer itself is not of interest, all the layer does is introduce effective slip for 
the bulk system which can be simulated using the slip parameter $p$.

Assuming the velocity profile of the fluid to be of the form
\begin{equation}
 u(z)=\left\{\begin{array}{ll}
                \sum_{n=0}^N a_n z^n & \quad 0\leq z\leq\epsilon\\
                \sum_{n=0}^N b_n z^n & \quad \epsilon<z\leq R\nonumber
            \end{array}\right.
\end{equation}
the slip length $u(0)/u'(0)$ can be determined analytically for $\epsilon\ll R$ 
where $R$ is the total system size and $\epsilon$ the thickness of the 
low-viscosity layer. Various conditions on $u$ give
\begin{eqnarray*}
 u(0)=0 & \Rightarrow & a_0 = 0\quad\textrm{no slip at wall}\\
 u(\epsilon_+) = u(\epsilon_-) & \Rightarrow & b_0 = \sum_{n=1}^N
\left(a_n-b_n\right)\epsilon^n=(a_1-b_1)\epsilon+\mathcal{O}(\epsilon^2)
\quad\textrm{continuity of }u\\
\eta_{II} u'(\epsilon_+) = \eta_{I}u'(\epsilon_-) & \Rightarrow & a_1 = 
\frac{\eta_{II}}{\eta_I}b_1+\mathcal{O}(\epsilon)\quad\textrm{continuity of stress}
\end{eqnarray*}
and therefore
\begin{equation}
 b_0 = b_1\left(\frac{\eta_{II}}{\eta_I}-1\right)\epsilon+\mathcal{O}(\epsilon^2)
\nonumber
\end{equation}

Observing that $u(0)=b_0$ and $u'(0)=b_1$ we arrive at
\begin{equation}
 \frac{u(0)}{u'(0)}=\epsilon\left(\frac{\eta_{II}}{\eta_I}-1\right)
\label{eq:uoverup2}
\end{equation}
to first order in $\epsilon$, where $\eta_{I}$ is the (lower) viscosity of the thin
layer close to the wall and $\eta_{II}$ the (higher) viscosity of the bulk system.

\begin{figure}[ht!]
    \begin{center}
 \includegraphics[width=0.7\textwidth]{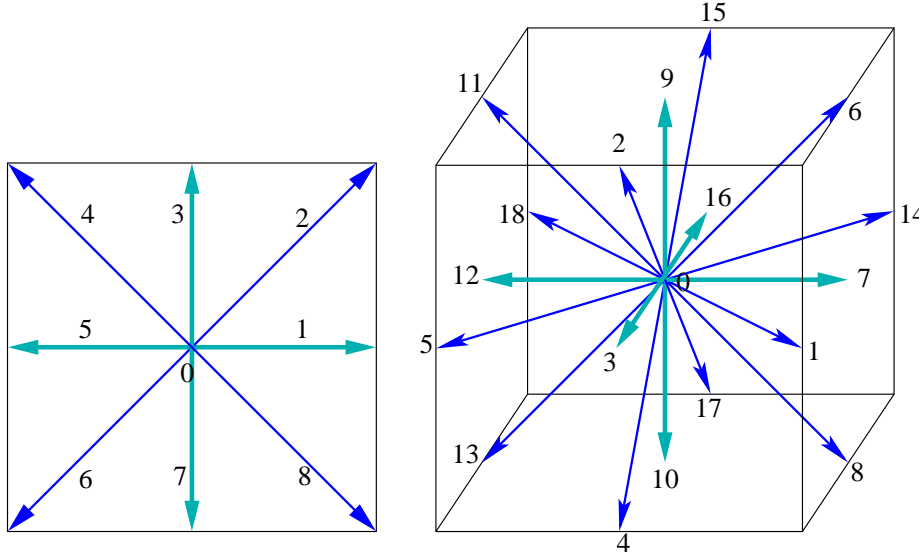}
    \end{center}
 \caption{(Online version in colour.) {\it Left}: Set of two-dimensional velocities $\mathbf{e}_i$ with 
	$i=0$ to 8, D2Q9, used in the following derivations. {\it Right}:
	Set of three-dimensional velocities $\mathbf{e}_i$ with $i=0$ to 18,
	D3Q19, used in the simulations.}
 \label{fig:lbvelocities}
\end{figure}

In order to relate slip length $u(0)/u'(0)$ and slip parameter $p$ and 
relaxation time $\tau$ in the LB method, we consider the D2Q9 model. All 
simulations have been performed in the D3Q19 model where the derivation works 
in exactly the same way but the algebra is more cumbersome (Fig.~\ref{fig:lbvelocities}).

In a collision step, velocity populations of the same node collide and relax 
towards their local and instantaneous equilibrium population 
$f_{i}^\mathrm{eq}(t,\mathbf{x})$
\begin{eqnarray}
 f_i(t,\mathbf{x}) & = & f_i^*(t,\mathbf{x}) - \frac{1}{\tau}\left(
		f_i^*(t,\mathbf{x})-f_{i}^\mathrm{eq}(t,\mathbf{x})\right)\nonumber\\ 
	  & = & f_i^*(t,\mathbf{x})\left(1-\frac{1}{\tau}\right)+\frac{1}{\tau}
		f_i^\mathrm{eq}(t,\mathbf{x})
 \label{eq:collision}
\end{eqnarray}
Here $f_i(t,\mathbf{x})$ is the velocity population of direction $i$ at time 
$t$ and node $\mathbf{x}$ after collision and $f_i^*(t,\mathbf{x})$ the
population before collision. The relaxation time $\tau$ determines how far the
population relaxes in each step and is related to viscosity by 
$\tau=\frac{1+6\eta_{II}}{2}$. Here only the bulk viscosity $\eta_{II}$ enters 
the formula as it is the bulk fluid that is simulated in LB whereas the 
low-viscosity layer  and thus $\eta_I$ is incorporated into the boundary 
condition.

The equilibrium populations are given by
\begin{equation}
 f^\mathrm{eq}_i(z) = \rho \omega_{i} \left(1+3\mathbf{e}_i\cdot\mathbf{u}(t,\mathbf{x})
	+ \frac{9}{2}\left(\mathbf{e}_i\cdot\mathbf{u}(t,\mathbf{x})\right)^2
	- \frac{3}{2}\mathbf{u(t,\mathbf{x})}^2\right)
\label{eq:f(u)}
\end{equation}
with $\omega_1=\omega_3=\omega_5=\omega_7=1/9=:w_1$ and 
$\omega_2=\omega_4=\omega_6=\omega_8=1/36=:w_2$. Density $\rho$ is assumed to be
constant and fluid velocity is given by
\begin{equation}
\rho\mathbf{u}(t,\mathbf{x})=\sum_i\mathbf{e}_i f_i(t,\mathbf{x}).
\end{equation}

We next need to make a couple of assumptions on the velocity profile. First, 
we assume that $\mathbf{u}(t,\mathbf{x})=u(t,z)\mathbf{e}_x$ is 
parallel to the wall and that it only depends on the distance to the wall $z$.
So for the velocity in $x$-direction we have
\begin{equation}
 \rho u(z)  = f^*_1(z)+f^*_2(z)-f^*_4(z) -f^*_5(z)-f^*_6(z)+f^*_8(z) 
 \label{eq:vel}
\end{equation}

In the streaming step velocity populations from time step $t$ are passed on to
neighbouring nodes and become pre-collision populations of time step $t+1$.
Below, streaming is given for the layer of fluid nodes just above the wall, i.e. 
those experiencing reflection from the wall. As velocities are assumed to be 
homogeneous in $x$- and $y$-directions those indices are omitted 
altogether and $f_{i,k}$ denotes velocity population $i$ in layer $k$ above the 
first fluid layer. For the first layer $k=0$ will be omitted.
\begin{eqnarray}
 f_1^*(t+1) & = & f_1(t)\nonumber\\
 f_2^*(t+1) & = & q f_6(t) + p f_8(t)\nonumber\\
 f_3^*(t+1) & = & f_7(t)\nonumber\\
 f_4^*(t+1) & = & p f_6(t) + q f_8(t)\nonumber\\
 f_5^*(t+1) & = & f_5(t)\nonumber\\
 f_6^*(t+1) & = & f_{6,1}(t)\nonumber\\
 f_7^*(t+1) & = & f_{7,1}(t)\nonumber\\
 f_8^*(t+1) & = & f_{8,1}(t)
 \label{eq:streaming}
\end{eqnarray}
The next assumption to be made is that steady-state has been reached and the 
velocity profile and all populations do no longer change with time meaning that
all reference to $t$ can be dropped.

Plugging streaming (Eq.~\ref{eq:streaming}) and collision (Eq.~\ref{eq:collision}) 
into Eq.~\ref{eq:vel} yields
\begin{eqnarray*}
 \rho u & = & f_1 + qf_6 + pf_8 - pf_6 - qf_8 - f_5 - f_{6,1} + f_{8,1}\\
	& = & f_1^*\left(1-\frac{1}{\tau}\right) + \frac{1}{\tau}f_1^\mathrm{eq}
		+\left(1-2p\right)\left(f_6^*\left(1-\frac{1}{\tau}\right) + 
		\frac{1}{\tau}f_6^\mathrm{eq}\right)\\
	& &	+\left(2p-1\right)\left(f_8^*\left(1-\frac{1}{\tau}\right) + 
		\frac{1}{\tau}f_8^\mathrm{eq}\right)
		- f_5^*\left(1-\frac{1}{\tau}\right) - \frac{1}{\tau}f_5^\mathrm{eq}\\
	& &	- f_{6,1}^*\left(1-\frac{1}{\tau}\right) - \frac{1}{\tau}f_{6,1}^\mathrm{eq}
		+ f_{8,1}^*\left(1-\frac{1}{\tau}\right) + \frac{1}{\tau}f_{8,1}^\mathrm{eq}
\end{eqnarray*}
To close this equation we need to express the pre-collision populations $f_i^*$
in terms of equilibrium populations $f_i^\mathrm{eq}$ as for those we have
Eq.~\ref{eq:f(u)} relating them to $\rho$ and $u$.

For populations $f_1^*$ and $f_5^*$ this is easy as no populations from higher
layers (larger $z$) are involved and the profile is assumed to be homogeneous in
$x$ meaning that after many iterations $f_1^*$ and $f_5^*$ simply relax to 
$f_1^\mathrm{eq}$ and $f_5^\mathrm{eq}$. For $f_6^*$ and $f_8^*$ this is more
complicated as populations from higher and higher layers will be involved
\begin{equation}
 f_6^*(t+n) = f_{6,n}^*(t) \left(1-\frac{1}{\tau}\right)^n
		+ \frac{1}{\tau} \sum_{k=1}^n 
			\left(1-\frac{1}{\tau}\right)^{k-1} f_{6,k}^\mathrm{eq}.
\end{equation}
For large enough $n$ the term involving $\left(1-\frac{1}{\tau}\right)^n$ can be
disregarded and using the above expression and Eq.~\ref{eq:f(u)} in the equation 
for $\rho u$ leaves us with
\begin{eqnarray*}
 \rho u_0 & = & f_1^\mathrm{eq} + \left(2p-1\right) \left(\frac{1}{\tau}\sum_{k=0}^n
 		\left(1-\frac{1}{\tau}\right)^k \left(f_{8,k}^\mathrm{eq}
  			- f_{6,k}^\mathrm{eq}\right)\right)\\
	& &	+ \frac{1}{\tau}\sum_{k=0}^n
 		\left(1-\frac{1}{\tau}\right)^k \left(f_{8,k+1}^\mathrm{eq}
 			- f_{6,k+1}^\mathrm{eq}\right)
 		- f_5^\mathrm{eq}\\
 	& = & \rho w_1 6u_0 + \frac{1}{\tau} w_2 6 \sum_{k=0}^n
 		\left(1-\frac{1}{\tau}\right)^k \left(2p u_k + u_{k+1} - u_k\right)
\end{eqnarray*}
where $u_0$ is the velocity in the layer of fluid nodes closest to the wall and
$u_k$ the velocities in higher layers.

Using Taylor expansions $u_k=\sum_{j=1}^\infty\frac{k^j}{j!}u_0^{(j)}$ for 
velocities at higher layers and combining subsequent terms of $u_k$ and 
$u_{k+1}$ in the sum above finally results in
\begin{equation}
 u_0 = w_1 6u_0 + \frac{1}{\tau} w_2 6 \left(\left(2p + \frac{1}{\tau-1}\right)
		\sum_{j=1}^\infty\frac{u_0^{(j)}}{j!} \sum_{k=0}^n
		k^j\left(1-\frac{1}{\tau}\right)^k -u_0 \frac{\tau}{\tau-1}\right)
\end{equation}
In principle $\sum_{k=0}^n k^j\left(1-\frac{1}{\tau}\right)^k$ can be evaluated
for arbitrary $j$ but then more and more derivatives $u_0^{(j)}$ have to be 
known in advance. We therefore truncated after the linear term and checked that
the resulting formula still holds within our numerical precision for Poiseuille
flow where the velocity profile in the channel is quadratic.

In the linear approximation $u_0^{(j)}=0$ for $j>1$ we obtain
 \begin{eqnarray*}
  u_0 & = & w_1 6 u_0 + \frac{1}{\tau} w_2 6 \left(\tau 2 p u_0 + 
 		\left(\tau^2-\tau\right)2p u_0' + \tau u_0'\right)\\
 	& = &	\frac{2}{3}u_0 + \frac{1}{3}p u_0 + \frac{1}{6}u_0' 
 		+ \frac{1}{3}\left(\tau-1\right)p u_0'.
 \end{eqnarray*}
At this point it is important to note that in line with standard LB
algorithms working within the Euler scheme which we use, the wall is
implemented between two nodes, so that it is mid-grid. Therefore $u(0)$
from Eq.~\ref{eq:uoverup} does not correspond to the velocity at the first layer
of fluid nodes $u_0$ but lies half a lattice spacing below $u(0)=u_0-1/2u_0'$.
As $u_0'$ is assumed constant it is of course the same as $u'(0)$ and we get
\begin{equation}
 \frac{u(0)}{u'(0)}=\frac{2\tau-1}{2}\frac{p}{1-p}
\end{equation}
which can be used together with Eq.~\ref{eq:uoverup2} to obtain the formula 
relating slip parameter $p$ and physical quantities $\epsilon$, $\eta_I$ and 
$\eta_{II}$ in Eq.~\ref{eq:pexplicit} of the {\it Methods and models} section.

}


\begin{thebibliography}{18}
\providecommand{\url}[1]{\texttt{#1}}
\providecommand{\urlprefix}{ }

\bibitem[Squires(2010)]{squires}
Squires, T.~M. 2010.
\newblock A furtive stare at an intra-cellular flow.
\newblock \emph{J. Fluid Mech.} \textbf{642}, 1--4.

\bibitem[Shimmen(2007)]{shimmen_sliding_2007}
Shimmen, T. 2007.
\newblock The sliding theory of cytoplasmic streaming: fifty years of progress.
\newblock \emph{J. Plant. Res.} \textbf{120}, 31--43.

\bibitem[Shimmen and Yokota(2004)]{shimmen_cytoplasmic_2004}
Shimmen, T. and Yokota, E. 2004.
\newblock Cytoplasmic streaming in plants.
\newblock \emph{Curr. Opin. Cell Biol.} \textbf{16}, 68--72.

\bibitem[Niwayama et~al.(2011)Niwayama, Shinohara, and
  Kimura]{niwayama_pnas_2011}
Niwayama, R., Shinohara, J., and Kimura, A. 2011.
\newblock Hydrodynamic property of the cytoplasm is sufficient to mediate
  cytoplasmic streaming in the {\em caenorhabiditis elegans} embryo.
\newblock \emph{Proc. Natl. Acad. Sci. USA} \textbf{108}, 11900--11905.

\bibitem[van()]{van_de_meent_natures_2008}
van~de Meent, J.-W., Tuval, I., and Goldstein, R.~E 2008.
\newblock Nature's microfluidic transporter: Rotational cytoplasmic streaming at high
  {P}eclet numbers.
\newblock \emph{Phys. Rev. Lett.} \textbf{101}, 178102--4.

\bibitem[Goldstein et~al.(2008)Goldstein, Tuval, and van~de
  Meent]{van_de_meent_pnas_2008}
Goldstein, R.~E., Tuval, I., and van~de Meent, J.-W. 2008.
\newblock Microfluidics of cytoplasmic streaming and its implications for
  intracellular transport.
\newblock \emph{Proc. Natl. Acad. Sci. USA} \textbf{105}, 3663--3667.

\bibitem[van~de Meent et~al.(2010)van~de Meent, Sederman, Gladden, and
  Goldstein]{van_de_meent_measurement_2010}
van~de Meent, J.-W., Sederman, A.~J.,  Gladden, L.~F., and Goldstein, R.~E. 2010.
\newblock Measurement of cytoplasmic streaming in single plant cells by
  magnetic resonance velocimetry.
\newblock \emph{J. Fluid Mech} \textbf{642}, 5--14.

\bibitem[Mustacich and Ware(1974)]{mustacich_observation_1974}
Mustacich, R.~V. and Ware, B.~R. 1974.
\newblock Observation of protoplasmic streaming by laser-light scattering.
\newblock \emph{Phys. Rev. Lett.} \textbf{33}, 617--620.

\bibitem[Mustacich and Ware(1976)]{mustacich_study_1976}
Mustacich, R.~V. and Ware, B.~R. 1976.
\newblock A study of protoplasmic streaming in nitella by laser doppler
  spectroscopy.
\newblock \emph{Biophys. J.} \textbf{16}, 373--388.

\bibitem[Sattelle and Buchan(1976)]{sattelle_cytoplasmic_1976}
Sattelle, D.~B. and Buchan, P. 1976.
\newblock Cytoplasmic streaming in chara corallina studied by laser light
  scattering.
\newblock \emph{J. Cell Sci.} \textbf{22}, 633--643.

\bibitem[Nothnagel and Webb(1982)]{nothnagel_hydrodynamic_1982}
Nothnagel, E.~A., and Webb, W.~W. 1982.
\newblock Hydrodynamic models of viscous coupling between motile myosin and
  endoplasm in characean algae.
\newblock \emph{J. Cell. Biol.} \textbf{94}, 444--454.

\bibitem[Kachar and Reese(1988)]{kachar_mechanism_1988}
Kachar, B., and Reese, T.~S. 1988.
\newblock The mechanism of cytoplasmic streaming in characean algal cells:
  sliding of endoplasmic reticulum along actin filaments.
\newblock \emph{J. Cell Biol.} \textbf{106}, 1545--1552.

\bibitem[Yamamoto et~al.(2006)Yamamoto, Shimada, Ito, Hamada, Ishijima,
  Tsuchiya, and Tazaw]{yamamoto_chara_2006}
Yamamoto, K., Shimada, K., Ito, K., Hamada, S., Ishijima, A., Tsuchiya, T., and
  Tazaw, M. 2006.
\newblock Chara myosin and the energy of cytoplasmic streaming.
\newblock \emph{Plant Cell Physiol.} \textbf{47}, 1427--1431.

\bibitem[Morimatsu et~al.(2000)Morimatsu, Nakamura, Sumiyoshi, Sakaba,
  Taniguchi, Kohama, and Higashi-Fujime]{morimatsu_molecular_2000}
Morimatsu, M., Nakamura, A., Sumiyoshi, H., Sakaba, N., Taniguchi, H., Kohama, K.,
  and Higashi-Fujime, S. 2000.
\newblock The molecular structure of the fastest myosin from green algae,
  {C}hara.
\newblock \emph{Biochem. Biophys. Res. Commun.} \textbf{270}, 147--152.

\bibitem[Cates et~al.(2004)Cates, Stratford, Adhikari, Stansell, Desplat,
  Pagonabarraga, and Wagner]{ludwig}
Cates, M.~E., Stratford, K., Adhikari, R., Stansell, P., Desplat, J.~C.,
  Pagonabarraga, I., and Wagner, A.~J. 2004.
\newblock Simulating colloid hydrodynamics with lattice boltzmann methods.
\newblock \emph{J. Phys. Condens. Matt.} \textbf{16}, S3903--S3915.

\bibitem[Nguyen and Ladd(2002)]{ladd}
Nguyen, N.~Q. and Ladd, A.~J.~C. 2002.
\newblock Lubrication corrections for lattice-boltzmann simulations of particle
  suspensions.
\newblock \emph{Phys. Rev. E} \textbf{66}, 046708.

\bibitem[Sbragaglia and Succi(2005)]{sbragaglia_analytical_2005}
Sbragaglia, M. and Succi, S. 2005.
\newblock Analytical calculation of slip flow in lattice boltzmann models with
  kinetic boundary conditions.
\newblock \emph{Phys. Fluids} \textbf{17}, 093602.

\bibitem{oliver}
Henrich, O., Stratford, K., Marenduzzo, D. and Cates, M.~E. 2010.
Ordering dynamics of blue phases entails kinetic stabilization of amorphous networks. {\it Proc. Natl. Acad. Sci. USA} {\bf 107}, 13212--13215.


\bibitem[Nagai and Rebhun(1966)]{nagai_cytoplasmic_1966}
Nagai, R. and Rebhun, L.~I. 1966.
\newblock Cytoplasmic microfilaments in streaming nitella cells.
\newblock \emph{J. Ultrastruct Res.} \textbf{14}, 571--89.

\bibitem{Eisenriegler}
Eisenriegler, E. 1993.
Polymers Near Surfaces. World Scientific, Singapore.

\end{thebibliography}
\end{document}